\documentclass[pra,twocolumn,superscriptaddress]{revtex4-2}
\usepackage{multirow}
\usepackage{rotating}
\usepackage{booktabs,tabularray}
\usepackage{amsmath}
\usepackage{mathrsfs}
\usepackage[protrusion=true,expansion=true]{microtype}
\usepackage{times}
\usepackage{bm,xcolor,braket, tikz}
\usepackage{graphicx}
\usepackage[normalem]{ulem}
\renewcommand\sout{\bgroup\markoverwith{\textcolor{magenta}{\rule[0.5ex]{2pt}{2pt}}}\ULon}

\usepackage{hyperref}
\hypersetup{colorlinks,linkcolor={blue},citecolor={blue},urlcolor={blue}}
\usepackage{amssymb,physics}
\usepackage[utf8]{inputenc}
\usepackage[english]{babel}
\usepackage{bm}

\makeatletter
\def\graphicscale{\twocolumn@sw{0.3}{0.4}}
\def\graphicthreescale{\twocolumn@sw{0.3}{0.4}}

\newcommand{\be}{\begin{equation}}
\newcommand{\ee}{\end{equation}}
\newcommand{\bea}{\begin{align}}
\newcommand{\eea}{\end{align}}

\usepackage{float}

\usepackage{dsfont}
\usepackage{mathrsfs}
\usepackage{comment}

\begin{document}
\title{Effects of intertube dipole-dipole interactions in nearly integrable one-dimensional $^{162}$Dy gases}

\author{Yicheng Zhang}
\affiliation{Homer L. Dodge Department of Physics and Astronomy,\\
The University of Oklahoma, Norman, Oklahoma 73019, USA}
\affiliation{Center for Quantum Research and Technology, The University of Oklahoma, Norman, Oklahoma 73019, USA}
\author{Kangning Yang}
\affiliation{Department of Physics, Stanford University, Stanford, California 94305, USA}
\affiliation{E. L. Ginzton Laboratory, Stanford University, Stanford, California 94305, USA}
\author{Benjamin L. Lev}
\affiliation{Department of Physics, Stanford University, Stanford, California 94305, USA}
\affiliation{E. L. Ginzton Laboratory, Stanford University, Stanford, California 94305, USA}
\affiliation{Department of Applied Physics, Stanford University, Stanford, California 94305, USA}
\author{Marcos Rigol}
\affiliation{Department of Physics, The Pennsylvania State University, University Park, Pennsylvania 16802, USA}

\begin{abstract}
We study the effects of the intertube dipole-dipole interactions (DDI) in recent experiments with arrays of nearly integrable one-dimensional (1D) dipolar Bose gases of $^{162}$Dy atoms. An earlier theoretical modeling ignored those interactions, which we include here via a modification of the 1D confining potentials. We investigate the effects of the intertube DDI both during the state preparation and during the measurements of the rapidity distributions. We explore how the strength of the contact interactions and the magnetic-field angles modify the intertube DDI corrections. We find that those corrections slightly change both the properties of the equilibrium state and the rapidity measurements. Remarkably, however, the changes nearly cancel each other, resulting in measured rapidity distributions that are very close to those predicted in the absence of the intertube DDI. 
\end{abstract}

\maketitle

\section{Introduction}\label{sec:intro}

Unlike for generic quantum many-body systems, \-near-inte\-grability\- makes possible the existence of long-lived quasiparticles that emerge from interactions among particles. The distribution of the rapidities (the momenta of the quasiparticles) fully characterizes the equilibrium states of nearly integrable systems~\cite{korepin_bogoliubov_book_93}. Understanding the out-of-equilibrium dynamics of those systems is one of the current research frontiers in quantum many-body physics. Interesting behaviors have been explored such as generalized thermalization beyond traditional statistical mechanics paradigms~\cite{rigol2007relaxation, cazalilla_06, rigol2007relaxation2, calabrese_11, gramsch_12, fagotti_13, wouters_denardis_14, pozsgay_mestyan14, ilievski_denardis_15, calabrese_essler_review_16, essler_fagotti_review_16,vidmar2016generalized}, fast prethermal dynamics and slow thermalization~\cite{Neyenhuis2017Manybody, Tang2018tni, Mori2018Thermalization, Mallayya2019Prethermalization, Ueda2020Quantum_review, le2023observation, Zhang2025Timescales}, generalized hydrodynamics (GHD)~\cite{castro2016emergent, bertini2016transport, Schemmer2019Generalized, malvania_zhang_21, Bouchoule2022Generalized, Moller2021Extension, Yang2024Phantom, Doyon2025Generalized, horvath2025observing}, and anomalous transport~\cite{Ljubotina2017Spin, Ljubotina2019KPZ, Scheie2021Detection, Wei2022KPZ, Gopalakrishnan2024Superdiffusion}. Well-defined quasiparticles and their rapidity distributions lie at the heart of the theoretical understanding of those behaviors.

Nearly integrable systems have generated much theoretical interest because of their elegant mathematical description and they have been realized in state-of-the-art ultracold-atom experimental platforms~\cite{Bloch2008Manybody, cazalilla_citro_review_11}. A well-studied example is arrays of one-dimensional (1D) bosonic gases with contact interactions trapped in deep 2D optical lattices~\cite{Kinoshita2005Local, kinoshita2006quantum}. These systems are described by the Lieb-Liniger model~\cite{lieb_liniger_63, olshanii_98}, which is a paradigmatic integrable model. A recent experimental breakthrough employing a modified time-of-flight (TOF) expansion sequence made possible the measurement of the rapidity distributions in such arrays~\cite{wilson_malvania_20}, enabling the observation of the dynamical fermionization of the momentum distribution of Tonks-Girardeau gases~\cite{Rigol2005Fermionization}. Using this protocol, experiments have tested the applicability of GHD~\cite{malvania_zhang_21}, measured the rapidity distributions of dipolar 1D ground-state gases~\cite{li_zhang_23}, characterized the nonlinear response of quantum many-body scar states~\cite{Yang2024Phantom}, studied 1D anyonic correlations~\cite{dhar2024anyonization}, and found evidence for Bethe strings~\cite{horvath2025observing}. Probing the local rapidity distribution has also been achieved in atom-chip experiments by selectively expanding a portion of the 1D gas~\cite{Dubois2024Probing}.

In this work we focus on the theoretical description of recent experiments with an array of nearly integrable 1D dipolar Bose gases of $^{162}$Dy atoms~\cite{li_zhang_23}. The long-range dipole-dipole interaction (DDI) between magnetic atoms introduces new features that are not available in 1D Bose gases interacting with purely contact interactions. For example, it provides a tunable integrability breaking term, enabling the study of thermalization near integrability~\cite{Tang2018tni, biagetti_lebek_24}. In addition, a repulsive DDI can stabilize gases with attractive contact interactions and enable the preparation of metastable super-Tonks-Girardeau gas states, including those within the quantum many-body scar regime~\cite{Kao2021tpo, Chen2023usg, Yang2024Phantom}. The rapidity and momentum distributions of the equilibrium states in the regime with repulsive contact interactions were studied in Ref.~\cite{li_zhang_23}. Notably, the theoretical model was able to describe the main experimental observations by accounting for only the intratube DDI (the DDI between bosons in the same 1D gas) via a modification of the contact interaction. The intertube DDI was ignored. (This is the DDI between bosons in different 1D gases.)

In the experimental work of Ref.~\cite{li_zhang_23}, the first step in the equilibrium state preparation was loading the Bose-Einstein condensate (BEC) into a deep 2D optical lattice (see Sec.~\ref{sec:exp}). That step was carried out using a magnetic field oriented at an angle at which the intratube DDIs vanish (approximately 55$^\circ$ with respect to the 1D tube axis). The intertube DDIs, which were present during that experimental step, were neglected in the theoretical modeling. The main goal of this work is to incorporate the intertube DDI into the theoretical modeling of that step in the experiment and then to account for it in the other steps, including the expansion used to access the rapidity distributions. While we expect the effect of the intertube DDI to be small in each stage of the modeling, it is important to understand how those effects add up.

We treat the leading effect of the intertube DDI as a modification to the 1D trapping potentials. It depends on the atom distribution across the entire 2D array of 1D gases. During the equilibrium state preparation, the intertube DDI affects the atom number and the temperature of each 1D gas. Later, it also affects the dynamics of the 1D gases during the expansion used to measure the rapidity distribution. For sufficiently long expansion times, the intertube DDI energy is converted into kinetic energy, and the rapidity distribution is slightly modified from that in the equilibrium state. Remarkably, we find that the changes in the rapidity distribution introduced by the intertube DDI (to the equilibrium state and during the expansion dynamics) nearly cancel each other, resulting in a measured rapidity distribution that is very close to that predicted in the absence of the intertube DDI. Therefore, the DDI is unlikely to be the cause of the disagreement between the model and experimental results in Ref.~\cite{li_zhang_23}. Nonthermal effects, due to the near-integrability of the 1D gases, are the most probable cause; we leave this exploration to future studies.

The paper is organized as follows. In Sec.~\ref{sec:model} we review the experimental setup and the theoretical modeling used in Ref.~\cite{li_zhang_23}. We then discuss how we account for the leading effect of the intertube DDI via a modification of the 1D trapping potentials. In Sec.~\ref{sec:initialstate} we systematically study the effect of the intertube DDI during the equilibrium state preparation in the experiments. In Sec.~\ref{sec:expansion} we discuss how the intertube DDI modifies the (otherwise conserved) rapidity distributions during the 1D expansion. Our main findings are summarized in Sec.~\ref{sec:summary}.

\section{Experimental setup and its modeling}\label{sec:model}

We review the experimental setup used in Ref.~\cite{li_zhang_23} to create arrays of nearly integrable 1D dipolar Bose gases of $^{162}$Dy atoms and how the rapidity distribution measurements were carried out. We then discuss the theoretical modeling used in Ref.~\cite{li_zhang_23} and how in this work we calculate the leading corrections introduced by the intertube DDI.

\subsection{Experimental sequence}\label{sec:exp}

\begin{figure}[!bt]
    \includegraphics[width=0.95\columnwidth]{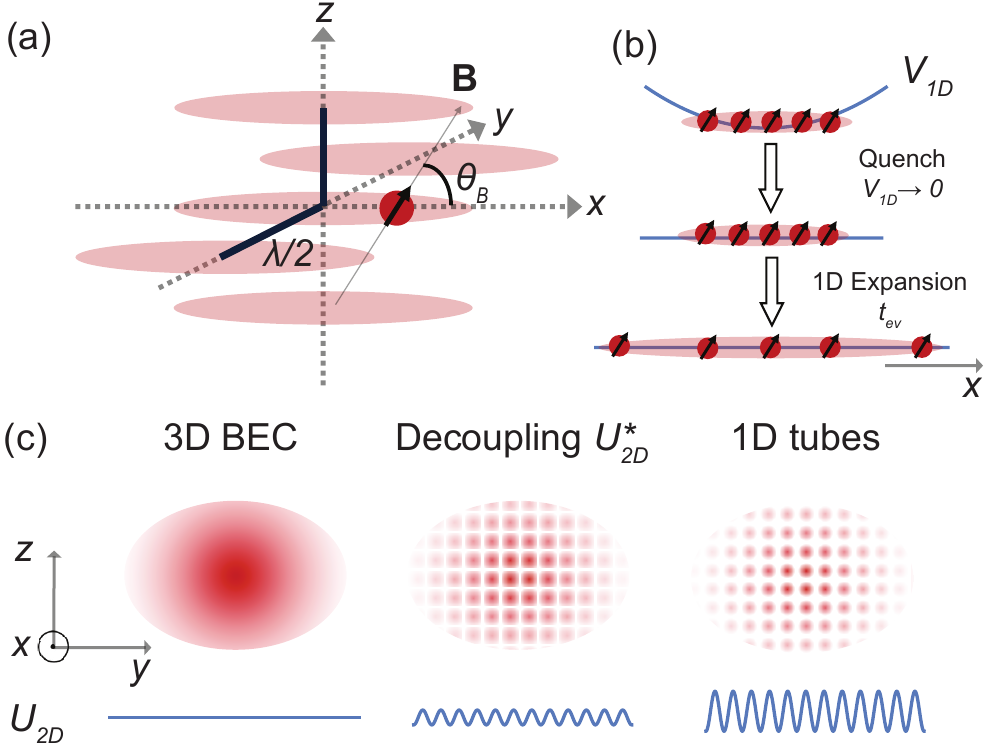}
    \caption{(a) Spatial depiction of the dipolar 1D gases. The 2D array (in the $y$-$z$ plane) of 1D gases (in the $x$ direction) is created using a deep 2D optical lattice with wavelength $\lambda$. The atomic dipoles align with the magnetic field $\bf B$ producing the intertube and intratube DDIs. (b) Sequence used to measure the rapidity distributions. After preparing the equilibrium 1D gases (tubes), the trapping potential $V_\mathrm{1D}$ is turned off and the atoms are allowed to expand in one dimension in the presence of interactions. The momentum distribution of the 1D gases becomes equal to the rapidity distribution after a long-enough expansion time $t_\text{ev}$. (c) Theoretical modeling of the experimentally created equilibrium state. Starting from a BEC (left), the 2D optical lattice potential $U_\mathrm{2D}$ is ramped up to create the array of tubes (right). We assume that at $U_\mathrm{2D}^*$ all the tubes decouple from each other, and that they are all at the same temperature $T_\mathrm{2D}^*$ (center). The ramping of 2D lattice is assumed to be adiabatic for $U_\mathrm{2D}>U_\mathrm{2D}^*$.}
    \label{fig:schematic}
\end{figure}

The experimental state preparation sequence in Ref.~\cite{li_zhang_23} begins with the creation of a $^{162}$Dy dipolar BEC via evaporative cooling in a crossed optical dipole trap (ODT). At the end of this step, the ODT frequencies are $(f_x,f_y,f_z)=(55.5,22.5,119.0)$ Hz. A typical BEC created in this setup contains $N_\text{tot}\sim2.3\times10^4$ atoms at a temperature of $38$~nK. The BEC is then loaded into a 2D optical lattice with wavelengh $\lambda=741$~nm to form an array of 1D gases (to which we also refer as tubes) [see Fig.~\ref{fig:schematic}(a)]. During loading, the depth of the 2D lattice $U_\mathrm{2D}$ is slowly (adiabatically) ramped up to 30$E_R$, creating strong confinements in the $y$ and $z$ directions. (We use the recoil energy $E_R=\hbar^2k_0^2/2m$ as the unit of energy, with $k_0=2\pi/\lambda$ and m the mass of a $^{162}$Dy atom.) In this experimental step, the angle $\theta_B$ of the magnetic field $\bf B$ in the $x$-$z$ plane [see Fig.~\ref{fig:schematic}(a)] is fixed at $\theta_B=55^{\circ}$ so that the intratube DDI vanishes.

Additional (adiabatic) experimental steps are then undertaken to prepare different equilibrium states and to make the rapidity measurements possible. A second ODT beam is turned on to cancel the antitrapping potential generated by the blue-detuned 2D lattice, and simultaneously the trapping frequency $f_x$ of the first ODT beam is reduced to 36.4~Hz. The magnetic field is then rotated to change the intratube DDI and/or the magnitude of the magnetic field is changed to control (via a Feshbach resonance) the strength of the contact interaction.

The rapidity measurement is done turning off the 1D trapping potential $V_\mathrm{1D}$ while keeping the 2D lattice $U_{2D}$ on and allowing the 1D gases to expand in the 1D direction for a time $t_\text{ev}$ [see Fig.~\ref{fig:schematic}(b)]. After a long expansion time $t_\text{ev}$, the momentum distribution of the 1D gases approaches the rapidity distribution~\cite{Rigol2005Fermionization, sutherland_98, minguzzi_gangardt_05, rigol_muramatsu_05c, campo_08, bolech_meisner_12}. A 3D TOF momentum measurement is then used to measure the asymptotic momentum distribution of all the atoms and therefore the rapidity distribution.

\subsection{Modeling uncoupled 1D gases}

The theoretical model used in Ref.~\cite{li_zhang_23} neglected the intertube DDI, i.e., it treated the experimental system as a 2D array of independent tubes. Within this approximation, each 1D gas can be described using the extended Lieb-Liniger model~\cite{lieb_liniger_63},
\begin{align}\label{eq:H_LL}
H&=\sum_{j=1}^{N}\left(-\frac{\hbar^2}{2m}\frac{\partial^2}{\partial x_j^2}+V_{\rm 1D}(x_j)\right)\nonumber\\&+\sum_{1\leq j<l \leq N}\left[g_{\rm 1D}^{\rm vdW}\delta(x_j-x_l)+U^{\rm intra}_{\rm DDI}(\theta_\text{B}, x_j-x_l)\right]\,,
\end{align}
which accounts for the confining potential $V_{\rm 1D}$ in the 1D direction and the intratube DDI $U^{\rm intra}_{\rm DDI}$ between the $N$ atoms. The effective 1D contact interaction $g_{\rm 1D}^{\rm vdW}$ is the result of the van der Waals force and depends on the depth of the 2D optical lattice $U_\text{2D}$ and the magnetic field (which determines the 3D $s$-wave scattering length $a_{\rm 3D}$). The explicit form of $g_{\rm 1D}^{\rm vdW}$ is
\begin{equation}\label{eq:g1d}
    g_{\rm 1D}^{\rm vdW}=-\frac{\hbar^2}{ma_{\rm 1D}}\,,
\end{equation}
where $a_{\rm 1D}=-a_{\perp}(a_{\perp}/a_{\rm 3D}-C)/2$ is the 1D scattering length,  $a_{\perp}=\sqrt{2\hbar/(m\omega_{\perp})}$ is the transverse confinement, $\omega_{\perp}=\sqrt{2U_{\rm 2D}k_0^2/m}$, and $C=-\zeta(1/2)\simeq1.4603$~\cite{olshanii_98}.

The effective intratube DDI $U^{\rm intra}_{\rm DDI}$, within the single mode approximation, is given by
\begin{equation}
U^{\rm intra}_{\rm DDI}(\theta_\text{B}, x_j-x_l)=\frac{\mu_0\mu^2}{4\pi}\frac{1-3\cos^2\theta_\text{B}}{\sqrt{2}a_{\perp}^3}\bigg(V^{\rm intra}_{\rm DDI}(u)-\frac{8}{3}\delta(u)\bigg)\,,
\end{equation}
where $V^{\rm intra}_{\rm DDI}(u)=-2|u|+\sqrt{2\pi}(1+u^2)e^{\frac{u^2}{2}}{\rm erfc}(\tfrac{|u|}{\sqrt{2}})$, $u=\tfrac{\sqrt{2}}{a_{\perp}}(x_j-x_l)$, ${\rm erfc}(x)$ is the complementary error function~\cite{Deuretzbacher2010gpo, Deuretzbacher2013egp, Tang2018tni, Kao2021tpo, DePalo2021pad}, and $\mu=9.93\mu_\text{B}$ is the   $^{162}$Dy atom dipole moment. 

In Ref.~\cite{li_zhang_23}, the 2D optical lattice used to created the array of 1D gases was ramped up such that $\theta_B=55^\circ$, which means that $U^{\rm intra}_{\rm DDI}\approx0$ during this process (studied in Sec.~\ref{sec:initialstate}) and in the ensuing measurements of the rapidity distribution (studied in Sec.~\ref{sec:expansion}). Also, $U^{\rm intra}_{\rm DDI}\approx0$ for initial states and their corresponding rapidity measurements in which different values of $g_{\rm 1D}^{\rm vdW}$ were explored experimentally using a Feshbach resonance at fixed $\theta_B=55^\circ$ (studied in Sec.~\ref{sec:sameangle}).

To account for the leading (short-range) effect of the intratube DDI when $\theta_B\neq 55^\circ$ (studied in Sec.~\ref{sec:diffangles}), we treat it as a modification to the contact interaction~\cite{li_zhang_23, Tang2018tni, Yang2024Phantom},
\begin{align}\label{eq:intraddi}
\tilde U^{\rm intra}_{\text{DDI}}(\theta_{B}, x_j-x_l)&=\frac{\mu_0\mu^2}{4\pi}\frac{1\!-\!3\cos^2\theta_{B}}{2a_{\perp}^2}\!\left(\!A-\frac{8}{3}\right)\!\delta(x_j\!-\!x_l)\nonumber\\ &\equiv g_{\rm 1D}^{\rm DDI}\delta(x_j\!-\!x_l)\,,
\end{align} 
where $A=\int_{-\infty}^{\infty}V^{\rm intra}_{\rm DDI}(u)du=4$~\cite{li_zhang_23}. Within this approximation, the Hamiltonian~\eqref{eq:H_LL} takes the form
\begin{equation}\label{eq:H_LL2}
\tilde H=\sum_{j=1}^{N}\bigg(-\frac{\hbar^2}{2m}\frac{\partial^2}{\partial x_j^2}+V_{\rm 1D}(x_j)\bigg)+\sum_{1\leq j<l \leq N}g_{\rm 1D}\delta(x_j-x_l)\,,
\end{equation}
where the effective 1D contact interaction is $g_{\rm 1D}=g_{\rm 1D}^{\rm vdW}+g_{\rm 1D}^{\rm DDI}$. In the absence of $V_{\rm 1D}$, Hamiltonian~\eqref{eq:H_LL2} is integrable and it is exactly solvable via the Bethe ansatz~\cite{lieb_liniger_63, yang_yang_69}. The ground-state observables depend only on the dimensionless parameter $\gamma=mg_\text{1D}/(\hbar^2 n_\text{1D})$, where $n_\text{1D}$ is the 1D density. When $V_{\rm 1D}$ is present, density and rapidity distributions can be obtained using a local density approximation (LDA).

\subsection{Modeling the state preparation in uncoupled 1D gases}

To describe the experimental results, in Ref.~\cite{li_zhang_23} we averaged the observables over all the tubes. We assumed that each tube is in thermal equilibrium after the state preparation, so the parameters that need to be determined, for each tube $\ell$ at position $(y_\ell, z_\ell)$, are the number of atoms $N_\ell$ and the temperature $T_\ell$. In Fig.~\ref{fig:schematic}(c) we illustrate the modeling and assumptions used to determine $N_\ell$ and $T_\ell$. 

Initially, the system is in a 3D BEC with total atom number $N_{\rm tot}$ and temperature $T_{\rm 3D}$ (left panel). As the 2D lattice is ramped up, the atoms are confined in coupled (number fluctuating) quasi-1D gases. We assumed that, at a lattice depth $U^*_{\rm 2D}$, the system decouples into independent 1D gases with fixed atom number (middle panel), and that all the tubes are in thermal equilibrium at the same temperature $T^*_{\rm 2D}$.

Under these assumptions and having knowledge of $N_{\rm tot}$ and the confining potentials in all directions, we used the thermodynamic Bethe ansatz~\cite{yang_yang_69} (TBA) (which describes 1D homogeneous Bose gases with contact interactions at finite temperature) and the LDA (to account for the confining potentials) to determine the atom number $N_\ell$ and entropy $S^*_\ell$ in each tube at the decoupling point for a dense grid of values of $U^*_{\rm 2D}$ and $T^*_{\rm 2D}$. The lattice depth $U^*_{\rm 2D}$ determines the contact interaction parameter $g^*_{\rm 1D}$ via Eq.~\eqref{eq:g1d}. We stress that, in the experiments, it is unlikely that all the tubes decouple at a single lattice depth $U^*_{\rm 2D}$. However, in Ref.~\cite{li_zhang_23} the theoretical results were found to be robust to changes in decoupling depth $U^*_{\rm 2D}$ near the optimal $U^*_{\rm 2D}$ and $T^*_{\rm 2D}$. (They are more sensitive to the value of $T^*_{\rm 2D}$.) Therefore, the assumption of a single decoupling depth is expected to be a reasonable approximation for our system.

Next we sequentially determined the temperatures of the 1D gases at the end of the optical lattice ramp [Fig.~\ref{fig:schematic}(c), right panel], after reducing the trapping frequency $f_x$ and after the magnetic field has been adjusted. To do so, we also use the TBA and LDA and assume that all the changes made after the decoupling depth $U^*_{\rm 2D}$ are adiabatic (i.e., that the entropy $S_\ell$ of each tube does not change) and that the 1D gases remain in thermal equilibrium throughout. Using the final temperatures and model parameters after all the experimental changes, we calculated the experimental observables. The two free parameters in our model, $U^*_{\rm 2D}$ and $T^*_{\rm 2D}$, were fixed by minimizing the difference between theoretical and experimental momentum and rapidity distributions.

\subsection{Modeling the effects of the intertube DDI}

In this work, except for the fact that we account for the effect of the intertube DDI, we follow the same assumptions as in Ref.~\cite{li_zhang_23}. To quantify the leading (energy changing) effect of the intertube DDI, we include it as a mean-field modification to the 1D trapping potentials,
\begin{equation}\label{eq:Vinter}
U_{\rm DDI}^{\rm inter}(x_\ell;y_\ell,z_\ell)=\sum_{\ell'\neq \ell}V_{\ell'}^{\rm inter}(x_\ell;y_\ell,z_\ell)\,,
\end{equation}
where
\begin{equation}
    V_{\ell'}^{\rm inter}(x_\ell;y_\ell,z_\ell)=\int dx_{\ell'} \frac{\mu_0\mu^2}{4\pi}\frac{1-3(\hat r\cdot \hat B )^2}{r^3}n(x_{\ell'};y_{\ell'},z_{\ell'})\,,
\end{equation}
$r=|{\bf r}|$, $\hat r=\tfrac{{\bf r}}{r}$, and ${\bf r}=(x_\ell-x_{\ell'},y_\ell-y_{\ell'}, z_\ell-z_{\ell'})$. 

Specifically, at the decoupling depth $U^*_{\rm 2D}$ determined in Ref.~\cite{li_zhang_23}, we use the temperature $T^*_{\rm 2D}$ and the density distribution computed in that work to calculate the modification introduced by Eq.~\eqref{eq:Vinter} to the trapping potentials. We then update the densities so that the system is in thermal equilibrium at the temperature $T^*_{\rm 2D}$ in the new trapping potentials. This procedure is repeated iteratively until the density distribution converges in all the tubes. We therefore end up with density and entropy distributions that differ from the ones in Ref.~\cite{li_zhang_23}. Similar iterative procedures, using the new density and entropy distributions, are repeated at each stage at which the temperatures of the 1D gases were determined in the absence of the intertube DDI in Ref.~\cite{li_zhang_23}. At the end of our new modeling of the equilibrium state preparation, we have slightly different density and temperature distributions when compared to those found in Ref.~\cite{li_zhang_23}. Finally, to model the rapidity measurements, we use GHD~\cite{castro2016emergent, bertini2016transport, Doyon2017Note} to study the expansion of the 1D gases in the weakening effective potential generated by the intertube DDI in the expanding gases.

\section{Initial state}\label{sec:initialstate}

To analyze in detail the effect of the intertube DDI on the initial state preparation for $\theta_B=55^{\circ}$ ($g_{\rm 1D}^{\rm DDI}=0$) and fixed $g_{\rm 1D}^{\rm vdW}$, we split our presentation into two sections. In Sec.~\ref{sec:decouple} we discuss the effect of the intertube DDI at the decoupling point and in Sec.~\ref{sec:finalstate} we discuss the effect of the intertube DDI through the end of the loading process.

\subsection{1D gases at the decoupling point $U^*_{\rm 2D}$}\label{sec:decouple}

In our calculations, we use the optimal values $U^*_{\rm 2D}=5E_R$ and $T^*_{\rm 2D}=25$~nK determined in Ref.~\cite{li_zhang_23}. Note that $T^*_{\rm 2D}=25$~nK is lower than $T_{3D}\sim 38$~nK, i.e., there is cooling when one transitions between three dimensions and one dimension~\cite{li_zhang_23} (see also Ref.~\cite{Guo2024Anomalous}). 

\begin{figure}[!t]
    \includegraphics[width=0.99\columnwidth]{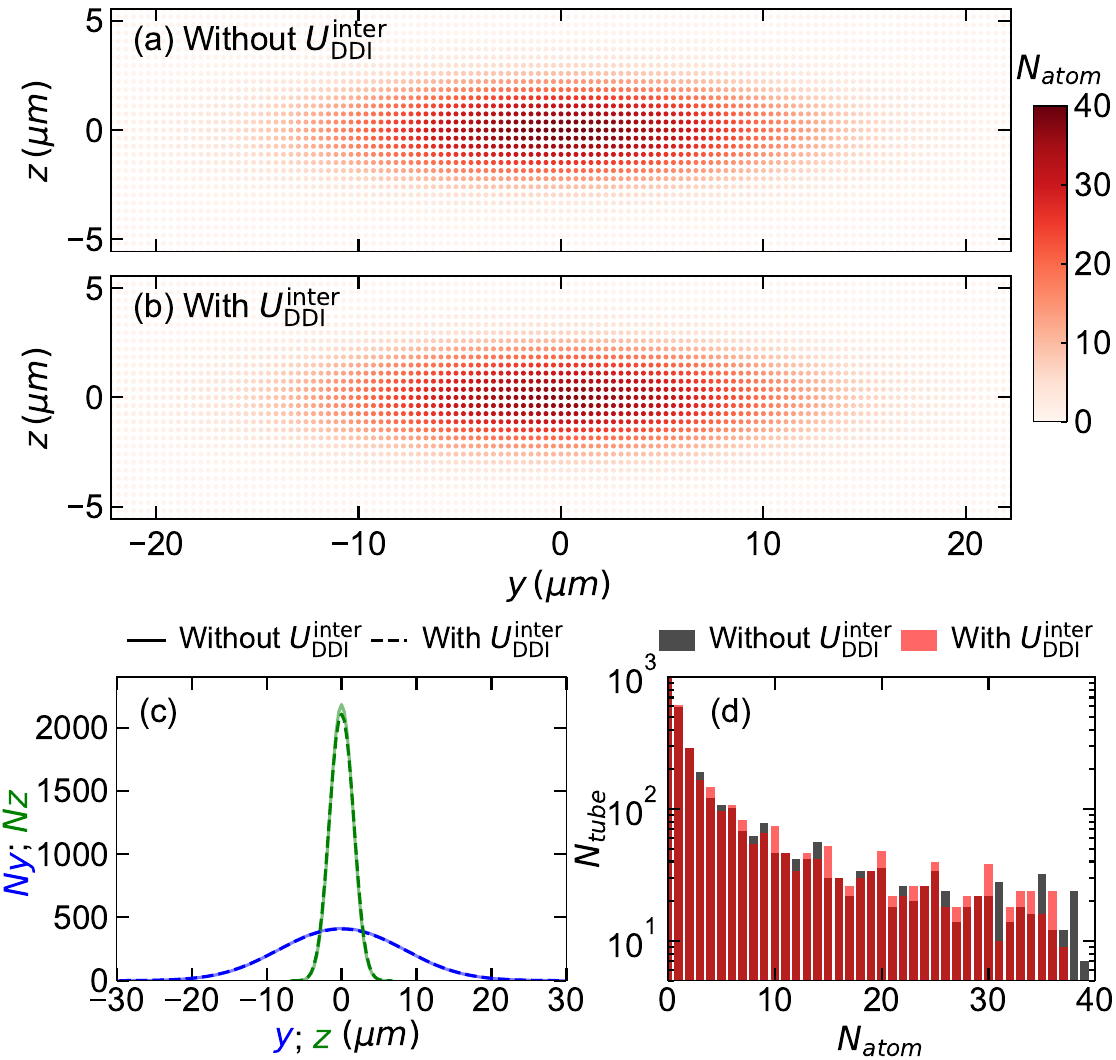}
    \vspace{-0.2cm}
    \caption{Atom distribution at decoupling. Distribution of atoms $N_\text{atom}(y_\ell,z_\ell)$ in the tubes when we (a) neglect and (b) account for the intertube DDI. Each dot indicates a tube $\ell$ at position $(y_\ell,z_\ell)$. (c) Distribution of atoms in the $y$ direction, $N_{y}=\sum_{z_\ell}N_\text{atom}(y_\ell,z_\ell)$ (blue lines), and in the $z$ direction, $N_{z}=\sum_{y_\ell}N_\text{atom}(y_\ell,z_\ell)$ (green lines), in calculations with (dashed) and without (solid) the intertube DDI correction. (d) Number of tubes $N_\text{tube}$ with $N_\text{atom}$ atoms vs $N_\text{atom}$.}
    \label{fig:atomdistribution}
\end{figure}

In Figs.~\ref{fig:atomdistribution}(a) and~\ref{fig:atomdistribution}(b) we show the atom distribution across all the tubes (dots) without and with the intertube DDI correction. The color indicates the number of atoms $N_\ell$ in each tube (see color scale on the right). The distributions are nearly indistinguishable from each other. In Fig.~\ref{fig:atomdistribution}(c) we plot the integrated atom distributions in the $y$ (blue lines) and $z$ (green lines) directions, which are also nearly indistinguishable from each other. The asymmetry in the size of the cloud in the different directions reflects the asymmetry of the employed trapping potential. 

To identify the nature of the differences between modeling with and without the intertube DDI, in Fig.~\ref{fig:atomdistribution}(d) we plot the number of tubes $N_{\rm tube}$ whose number of atoms is $N_{\rm atom}$ as a function of $N_{\rm atom}$. Since our modeling is done within the LDA, $N_\ell$ can take noninteger values in the tubes. To make the histogram shown in Fig.~\ref{fig:atomdistribution}(d), $N_\ell$ is rounded to the closest integer ($N_{\rm tube}$ for $N_{\rm atom}=0$ in the plot is the number of tubes with $0<N_\ell<0.5$). Figure~\ref{fig:atomdistribution}(d) shows that the intertube DDI decreases the maximal occupation of the tubes, i.e., in our asymmetric geometry [see Figs.~\ref{fig:atomdistribution}(a)--\ref{fig:atomdistribution}(c)] it acts like a weak antitrap in the $y$-$z$ plane.

\begin{figure}[!b]
    \includegraphics[width=0.99\columnwidth]{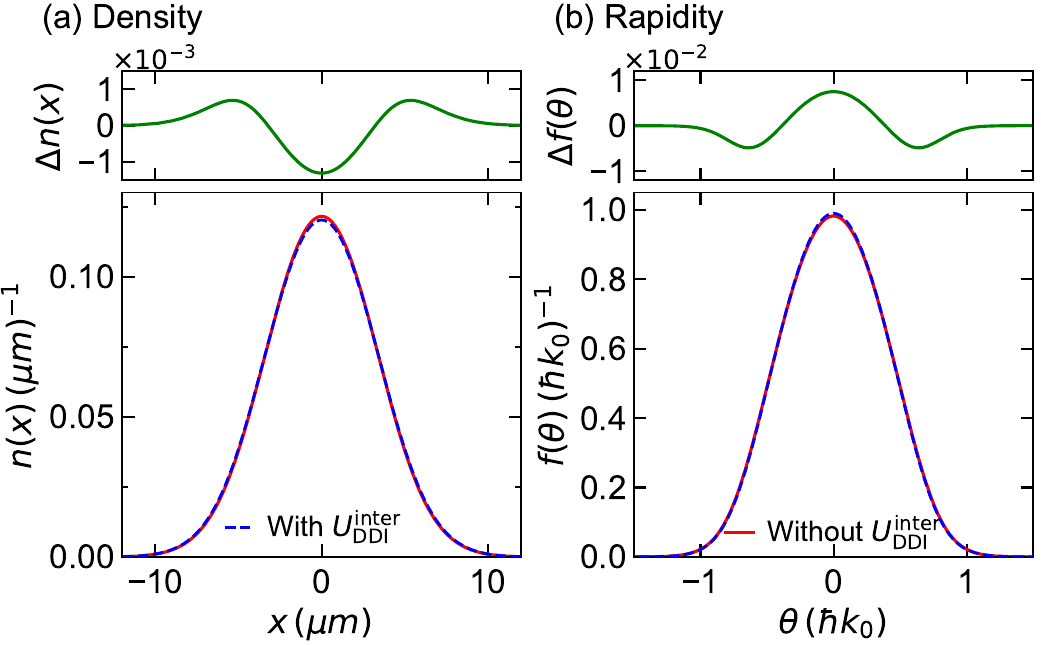}
    \vspace{-0.2cm}
    \caption{Density and rapidity distributions at decoupling. (a) Average density distribution of the 1D gases with (dashed lines) and without (solid lines) including the intertube DDI correction at decoupling. The green solid line in the top panel shows the difference $\Delta n(x)=n(x;{\rm with\,} U^{\rm inter}_{\rm DDI})-n(x;{\rm without\,} U^{\rm inter}_{\rm DDI})$. (b) Same as (a) but for rapidity distribution $f(\theta)$, with $\Delta f(\theta)=f(\theta; {\rm with\,} U^{\rm inter}_{\rm DDI})-f(\theta;{\rm without\,} U^{\rm inter}_{\rm DDI})$.}
    \label{fig:ddidecouple}
\end{figure}

The main observables of interest in this work, the averaged density $n(x)$ and the rapidity distribution $f(\theta)$, are shown at the decoupling point in the main panels of Figs.~\ref{fig:ddidecouple}(a) and~\ref{fig:ddidecouple}(b), respectively. The differences between the results without and with the intertube DDI are small. To highlight their nature, we magnify them at the top of each panel in Fig.~\ref{fig:ddidecouple}. They show that, in the presence of the intertube DDI, the density is slightly lower about the center of the trap (the density distribution is therefore slightly broader) and the occupation of the low rapidity modes is slightly higher (the rapidity distribution is slightly narrower). Hence, like in the $y$-$z$ plane, the intertube DDI in our asymmetric geometry acts like a weak antitrap in the $x$ direction.

\subsection{1D gases at the end of the loading process}\label{sec:finalstate}

A noticeable feature of Fig.~\ref{fig:atomdistribution}(d) obtained from finite-temperature modeling within the LDA is the large number of tubes containing less than two atoms. Despite their abundance, they only account for less than $5\%$ of the total number of atoms. Given the self-consistent nature of the calculations done in this work to account for the intertube DDI, including those tubes is computationally very demanding and they almost do not change the results: Their density is very low and they are located at the edge of the system; thus, they contribute very little to the total intertube DDI energy. Also, ignoring those tubes almost does not change the average rapidity distribution in the absence of the intertube DDI (see the Appendix). Therefore, we ignore them in the remainder of this paper, in both the calculations in which the intertube DDI is included and those in which it is not. We stress that, in contrast to the modeling in Ref.~\cite{li_zhang_23}, in this work we do not round $N_\ell$ to the closest integer in any of our calculations. In the Appendix we show that, in the absence of the intertube DDI, the rapidity distribution obtained with no rounding is closer to the experimental one than the one obtained in Ref.~\cite{li_zhang_23} with rounding, so here we remove this additional simplification of Ref.~\cite{li_zhang_23}. 

After the tubes decouple, we assume that the further increase in the depth of the 2D optical lattice, the canceling of the antitrapping potential, and the reduction of the trapping frequency in the $x$ direction ($f_x$: $55.5$ Hz$\to36.4$ Hz) take place adiabatically for each tube independently. (The global thermalizing effects of the intertube interactions occur in a much longer time scale~\cite{Tang2018tni}.) More specifically, we compute the final temperature $T_\ell$ of each tube at the end of the loading assuming that the entropy of each tube is the same as that at the decoupling point $S_\ell$. In practice, we search for $T_\ell$ (on a temperature grid with 0.5~nK steps) such that the entropy matches $S_\ell$ at the final $g_{\rm 1D}$ and 1D trapping potential including the intertube DDI modification. The intertube DDI correction introduced at the decoupling point affects this step through the modification of $N_\ell$ and $S_\ell$. In addition, the intertube DDI modifies the trapping potential, requiring an iterative calculation of $n_\ell(x)$ and $T_\ell$ until convergence is achieved. The effective contact interaction at the end of state preparation is $g^{\rm vdW}_{\rm 1D}=8.5$ $\frac{\hbar^2}{m}\,\mu \text{m}^{-1}$, and we find that the corresponding average effective dimensionless parameter is $\gamma_T=6.7$ (6.5) in the presence (absence) of the intertube DDI. This is nearly unchanged from $\gamma_T=6.7$ obtained in Ref.~\cite{li_zhang_23}. A motivation and detailed discussion of our definition of $\gamma_T$ in 2D arrays of inhomogeneous finite-temperature 1D gases can be found Ref.~\cite{li_zhang_23}. 

\begin{figure}[!t]
    \includegraphics[width=0.99\columnwidth]{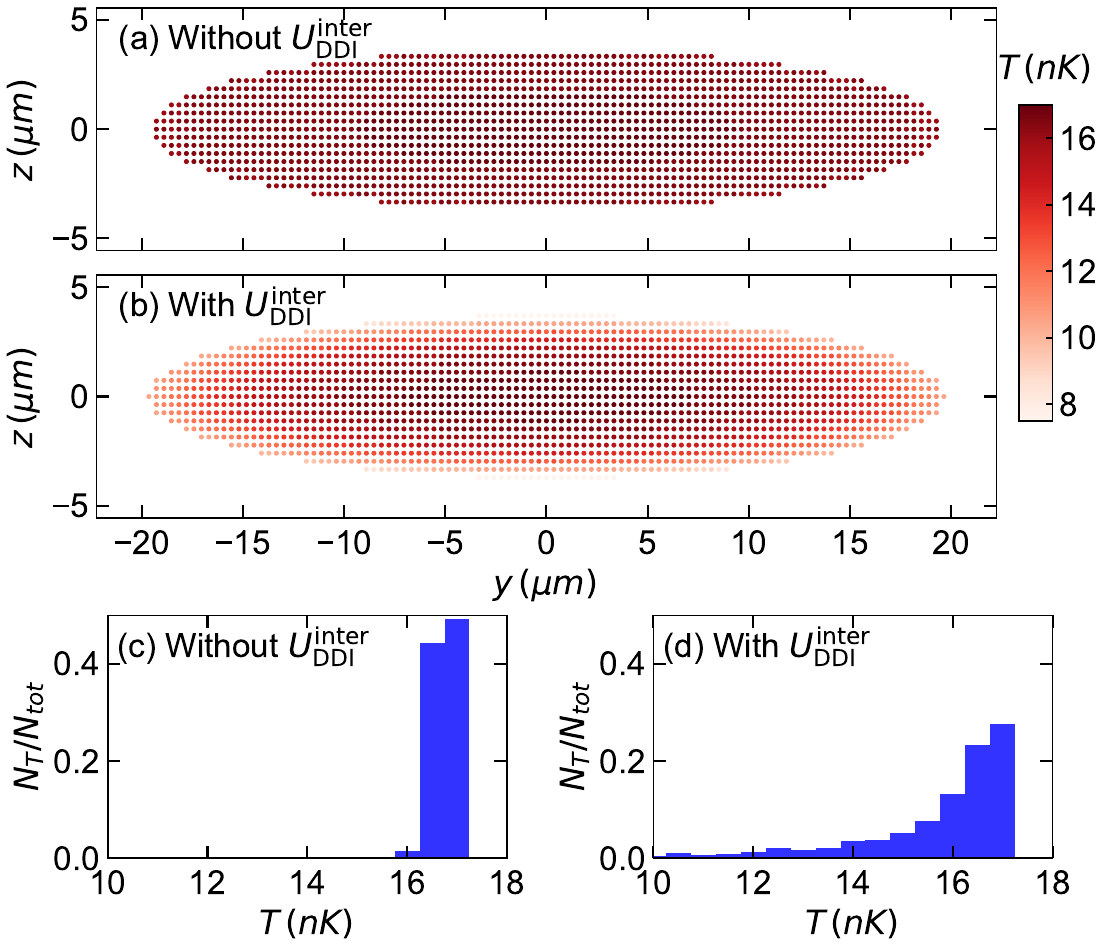}
    \vspace{-0.2cm}
    \caption{Temperature distribution at the end of the loading process. Temperature for tubes (a) without and (b) with intertube DDI after the state preparation. Each dot indicates a 1D tube at position $(y,z)$. Also shown is the ratio between atoms at temperature $T$ ($N_T$) and the total number of atom $N_{\rm tot}$ after the state preparation (c) without and (d) with intertube DDI correction. The temperature is binned in $0.5$-nK intervals. We show results for tubes with $N_\ell\geq 2$ atoms.}
    \label{fig:tubetemperature}
\end{figure}

Figures~\ref{fig:tubetemperature}(a) and~\ref{fig:tubetemperature}(b) show the temperature $T_\ell$ distribution without and with intertube DDI correction, respectively, for all tubes with $N_\ell\geq 2$.  Figures~\ref{fig:tubetemperature}(c) and~\ref{fig:tubetemperature}(d) show the corresponding distributions of the fraction of atoms $N_T/N_{\rm tot}$ with a temperature $T$ plotted as a function of $T$. Due to the reduction of trapping potential in the $x$ direction in the final stage of state preparation, the temperature of the tubes decreases significantly from the value at decoupling ($T^*=25$ nK) in both cases. In this final step of the loading process, we find that the intertube DDI correction has the clear effect of lowering the temperature of the tubes, specially of those that have lower filling (the ones away from the center of the system). This is again consistent with the intertube DDI acting as an antitrap in the $x$ direction, as we found at the decoupling point.

\begin{figure}[!t]
    \includegraphics[width=0.99\columnwidth]{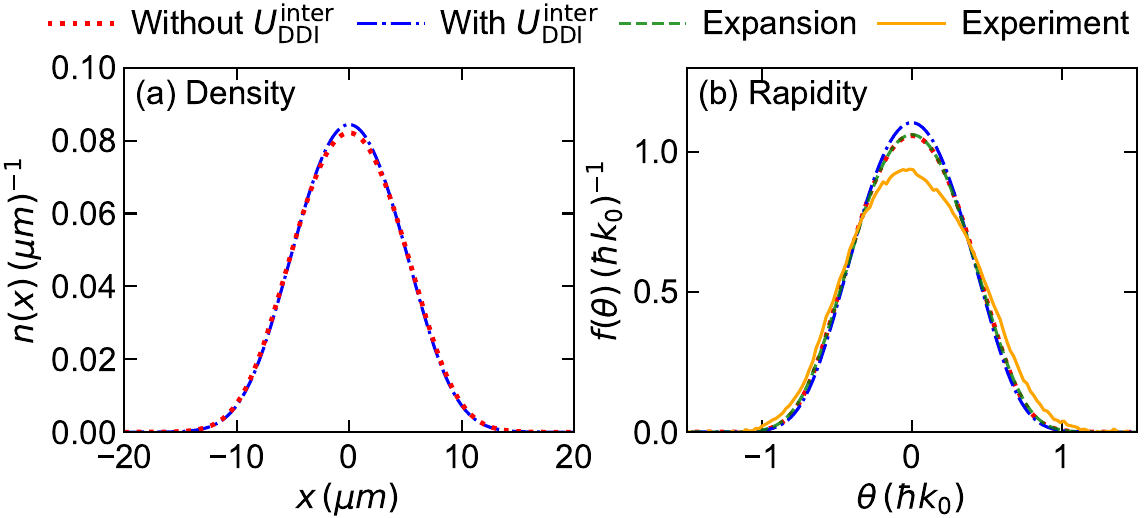}
    \vspace{-0.2cm}
    \caption{Density and rapidity distributions at the end of the loading process. (a) Density distributions $n(x)$ at the end of the loading process without (red dotted line) and with (blue dash-dotted line) the intertube DDI correction. (b) Rapidity distributions $f(\theta)$ at the end of the loading process without (red dotted line) and with (blue dash-dotted line) the intertube DDI correction. We also show the rapidity distribution obtained after a 15 ms expansion in the tubes (green dashed line, overlapping with the red dotted line), and the experimental results (orange solid line) from Ref.~\cite{li_zhang_23}.}
    \label{fig:DDIfinal}
\end{figure}

In Fig.~\ref{fig:DDIfinal}(a) we show the average density distributions $n(x)$ at the end of the state preparation without (red dotted line) and with (blue dashed dotted line) the intertube DDI correction. They are very close to each other despite the clearly different temperature distributions seen in Fig.~\ref{fig:tubetemperature}. This can be understood by considering that the lower temperature decreases the width of the density distribution while the effective antitrapping potential generated by the intertube DDI increases it. Those two effects compete with each other, resulting in a density distribution in the presence of the DDI that is very close to that in the absence of the DDI. In this particular case, the aforementioned competition results in a density distribution that is slightly narrower in the presence of the DDI than in its absence. 

In Fig.~\ref{fig:DDIfinal}(b) we show the corresponding average rapidity distributions. For the rapidity distribution, the intertube DDI correction to the temperatures (lowering the temperature) and the trapping potentials (antitrapping) both lower the width of the rapidity distribution. Hence, we find that the narrowing of the rapidity distribution due to the intertube DDI is more significant than that of the density distribution. In Fig.~\ref{fig:DDIfinal}(b) we also show the rapidity distribution measured in Ref.~\cite{li_zhang_23}. One can see that when the effect of the intertube DDI is included only in the state preparation (as we have done so far), the modeled rapidity distribution (slightly) differs more from the experimental one than that obtained in the absence of the DDI. As mentioned in Ref.~\cite{li_zhang_23}, one also needs to include the effect of the intertube DDI during the expansion, which we do next.

\section{Rapidity measurement}\label{sec:expansion}

To measure the rapidity distribution the 1D trapping potential is suddenly turned off ($V_{\rm 1D}\to0$) and the 1D gases are allowed to expand for a time $t_{\rm ev}$ while the 2D lattice $U_{\rm 2D}$ is still on. During the expansion, the atoms interact and their momentum distribution changes with time, asymptotically approaching the rapidity distribution at long times~\cite{wilson_malvania_20, Rigol2005Fermionization, sutherland_98, minguzzi_gangardt_05, campo_08, bolech_meisner_12}. In the absence of a DDI, i.e., for atoms interacting only via contact interactions, the rapidity distribution is conserved during the expansion. Therefore, measuring the momentum distribution after a long $t_{\rm ev}$ is equivalent to measuring the rapidity distribution of the initial state. In the presence of the DDI, the rapidity distribution changes during the expansion. Since the intertube DDI acts like an effective antitrapping potential, it accelerates the quasiparticles and broadens the rapidity distribution, namely, the effect of the intertube DDI on the rapidity distribution during the expansion dynamics is the opposite of the effect it has in the initial state.

We use GHD~\cite{castro2016emergent, bertini2016transport, Doyon2017Note} to compute the density $n(x;t_{\rm ev})$ and the rapidity distribution $f(\theta;t_{\rm ev})$ at $t_{\rm ev}=15$ ms, which is the 1D expansion time in the experiments before the 2D lattice is turned off and the momentum distribution of the atoms (which we assume is the same as the rapidity distribution at that time) is measured using a standard TOF expansion. In our simulation, all the 1D gases expand in a weakening antitrap generated by the DDI [see Eq.~\eqref{eq:Vinter}], which we update throughout the expansion. 

\begin{figure}[!t]
    \includegraphics[width=0.99\columnwidth]{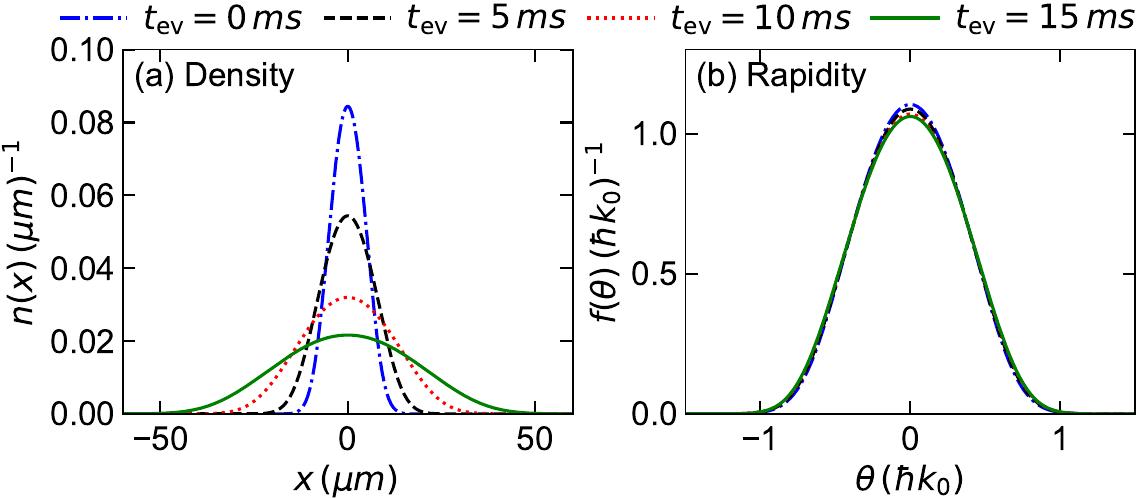}
    \vspace{-0.2cm}
    \caption{Density and rapidity distributions during the expansion. (a) Density distribution $n(x)$ at 1D expansion time $t_{\rm ev}$. We show results for $t_{\rm ev}=$ 0 ms (blue dash-dotted line), 5 ms (black dashed line), 10 ms (red dotted line), and 15 ms (green solid line). (b) Same as (a) but for the rapidity distribution $f(\theta)$.}
    \label{fig:DDIexpansion}
\end{figure}

In Fig.~\ref{fig:DDIexpansion}(a) we show the density distribution $n(x)$ at different times $t_{\rm ev}$ during the expansion in one dimension in the presence of the intertube DDI. After the 1D trap is turned off, the atom cloud expands ballistically and by $t_{\rm ev}=15$~ms it has a size that is about four times the initial size, as needed for the momentum distribution to evolve into the rapidity distribution~\cite{wilson_malvania_20}. By that time, we find that the remaining intertube DDI energy is about 1.5\% of the energy in the initial state, namely, almost all of the intertube DDI energy has been converted into kinetic energy. Figure~\ref{fig:DDIexpansion}(b) shows the corresponding evolution of the rapidity distribution. The changes in $f(\theta)$ occur mostly within expansion times $t_{\rm ev}\lesssim10$ ms. We find the rapidity distribution to be nearly stationary by $t_{\rm ev}=15$~ms and, as expected, broader than the initial rapidity distribution.

The rapidity distribution at $t_{\rm ev}=15$~ms is also shown in Fig.~\ref{fig:DDIfinal}(b) along with the other rapidity distributions discussed in Sec.~\ref{sec:finalstate}. Remarkably, it is almost indistinguishable from the rapidity distribution in the absence of the intertube DDI, but still significantly different from the experimental rapidity distribution. Our results suggest that the DDI is not responsible for the differences identified in Ref.~\cite{li_zhang_23}. Therefore, we expect that nonthermal effects due to the integrability of the 1D gases are responsible for those differences.

\section{Other initial states}\label{sec:otherIstates}

Next we explore how the modeling with and without the intertube DDI correction compare to one another for other initial states experimentally created in Ref.~\cite{li_zhang_23}. We consider states with $\theta_B=55^{\circ}$ ($g_{\rm 1D}^{\rm DDI}=0$) but different values of $g_{\rm 1D}^{\rm vdW}$ in Sec.~\ref{sec:sameangle}, and then states with different $\theta_B$ in Sec.~\ref{sec:diffangles}.

\subsection{Fixed $\theta_B=55^{\circ}$ and different $g_{\rm 1D}^{\rm vdW}$}\label{sec:sameangle}

We first consider the case in which after the experimental sequence studied in Sec.~\ref{sec:initialstate} is completed, the magnitude of the magnetic field is changed. As a result, the contact interaction $g^{\rm vdW}_{\rm 1D}$ changes via a Feshbach resonance. We consider the two experimental cases studied in Ref.~\cite{li_zhang_23}, with $g^{\rm vdW}_{\rm 1D}=4.1$ and $20.4$~$\frac{\hbar^2}{m}\,\mu \text{m}^{-1}$ resulting in a final effective $\gamma_T=3.1$ (3.1) and 17 (16), respectively, in the presence (absence) of the intertube DDI. Since $\theta_B$ is kept fixed at $\theta_B=55^\circ$, the intratube DDI remains near zero and the intertube DDI is not different from our previous example.

\begin{figure}[!t]
    \includegraphics[width=0.99\columnwidth]{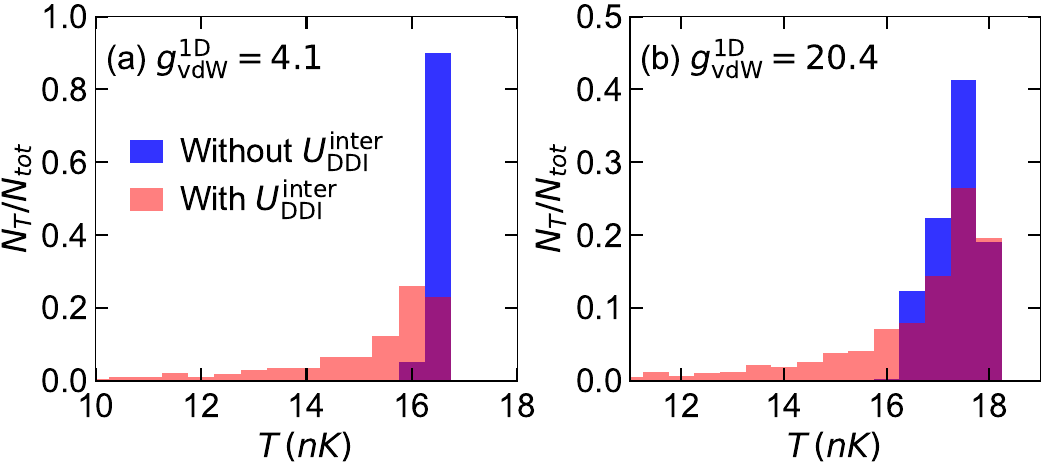}
    \vspace{-0.2cm}
    \caption{Temperature distribution for different values of $g_{\rm 1D}^{\rm vdW}$. Here $N_T/N_{\rm tot}$ is plotted for other initial states with $\theta_B=55^{\circ}$ and (a) $g^{\rm vdW}_{\rm 1D}=4.1$ $\frac{\hbar^2}{m}\,\mu \text{m}^{-1}$ and (b) $g^{\rm vdW}_{\rm 1D}=20.4$ $\frac{\hbar^2}{m}\,\mu \text{m}^{-1}$. Experimentally, $g^{\rm vdW}_{\rm 1D}$ is changed after loading into the deep 2D lattice by adiabatically modifying the magnitude of the magnetic field $B$ while keeping fixed its direction $\theta_B=55^{\circ}$. In our modeling, we modify $T$ in intervals of $0.5$~nK to find the closest entropy in each tube to that at decoupling.}
    \label{fig:temperaturegamma}
\end{figure}

\begin{figure}[!b]
    \includegraphics[width=0.99\columnwidth]{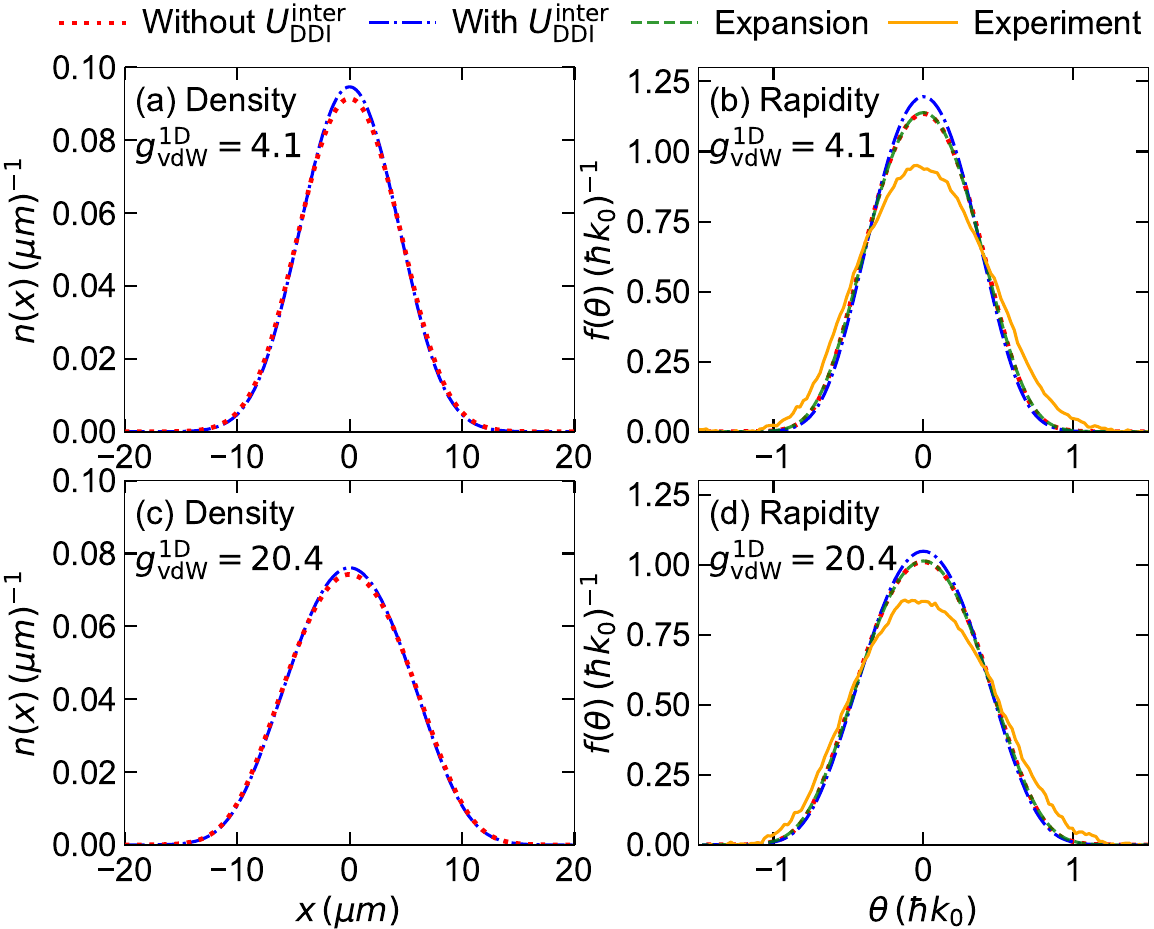}
    \vspace{-0.2cm}
    \caption{Density and rapidity distributions for different values of $g_{\rm 1D}^{\rm vdW}$. (a) and (c) Density and (b) and (d) rapidity distributions for (a) and (b) $g^{\rm vdW}_{\rm 1D}=4.1$ $\frac{\hbar^2}{m}\,\mu \text{m}^{-1}$ and (c) and (d) $20.4$ $\frac{\hbar^2}{m}\,\mu \text{m}^{-1}$. 
    }
    \label{fig:rapiditygamma}
\end{figure}

In Fig.~\ref{fig:temperaturegamma} we show our results for the distribution of temperatures of the states with $g^{\rm vdW}_{\rm 1D}=4.1$ $\frac{\hbar^2}{m}\,\mu \text{m}^{-1}$ [Fig.~\ref{fig:temperaturegamma}(a)] and $g^{\rm vdW}_{\rm 1D}=20.4$ $\frac{\hbar^2}{m}\,\mu \text{m}^{-1}$ [Fig.~\ref{fig:temperaturegamma}(b)]. Changing $g^{\rm vdW}_{\rm 1D}$ in the range of values experimentally explored does not significantly change the temperature of the tubes. The largest changes in the temperatures are due to the dimensional reduction from three dimensions to one dimension and due to lowering $f_x$ at the end of the loading process. We find that reducing $g^{\rm vdW}_{\rm 1D}$ slightly lowers the temperature of the tubes. As for the case discussed in Sec.~\ref{sec:finalstate} in which $g^{\rm vdW}_{\rm 1D}=8.5$ $\frac{\hbar^2}{m}\,\mu \text{m}^{-1}$, we also find that the intertube DDI slightly lowers the temperature and that its effect is more pronounced for the tubes with a lower number of atoms.

In Fig.~\ref{fig:rapiditygamma} we show the density $n(x)$ [Figs.~\ref{fig:rapiditygamma}(a) and~\ref{fig:rapiditygamma}(c)] and rapidity $f(\theta)$ [Figs.~\ref{fig:rapiditygamma}(b) and~\ref{fig:rapiditygamma}(d)] distributions for the states with $g^{\rm vdW}_{\rm 1D}=4.1$ $\frac{\hbar^2}{m}\,\mu \text{m}^{-1}$ [Figs.~\ref{fig:rapiditygamma}(a) and~\ref{fig:rapiditygamma}(b)] and $g^{\rm vdW}_{\rm 1D}=20.4$ $\frac{\hbar^2}{m}\,\mu \text{m}^{-1}$ [Figs.~\ref{fig:rapiditygamma}(c) and~\ref{fig:rapiditygamma}(d)]. As expected, reducing $g^{\rm vdW}_{\rm 1D}$ results in a narrowing of both the density and the rapidity distributions. The results are qualitatively similar to those for $g^{\rm vdW}_{\rm 1D}=8.5$ $\frac{\hbar^2}{m}\,\mu \text{m}^{-1}$ in Sec.~\ref{sec:finalstate}. With the intertube DDI correction, the density and rapidity distributions in the initial state are slightly narrower than in the absence of the correction and, after the expansion, the rapidity distribution with the DDI correction is nearly indistinguishable from that in the absence of the correction. As discussed in Ref.~\cite{li_zhang_23}, the rapidity distributions from our simulations approach the ones measured experimentally as $g^{\rm vdW}_{\rm 1D}$ ($\gamma_T$) increases.

\subsection{Fixed $g_{\rm 1D}^{\rm vdW}$ and different $\theta_B$}\label{sec:diffangles}

Finally, we consider the case in which after the experimental sequence studied in Sec.~\ref{sec:initialstate} is completed, the magnetic field is rotated without change in amplitude. The contact interaction $g^{\rm vdW}_{\rm 1D}=8.5$ $\frac{\hbar^2}{m}\,\mu \text{m}^{-1}$ does not change with the rotation of the magnetic field, but both the intratube (which becomes nonzero) and the intertube DDI change. To account for the intratube DDI, we use the approximation in Eq.~\eqref{eq:intraddi}, i.e., we treat it as a correction $g^{\rm DDI}_{\rm 1D}$ to the contact interaction. Three states with different $\theta_B\neq55^{\circ}$ were prepared in Ref.~\cite{li_zhang_23} using that sequence: (i) one with $\theta_B=90^\circ$ resulting in an additional effective repulsive contact interaction with $g^{\rm DDI}_{\rm 1D}=2.7$, (ii) one with $\theta_B=35^\circ$ resulting in an additional effective attractive contact interaction with $g^{\rm DDI}_{\rm 1D}=-2.7$, and (iii) one with $\theta_B=0^\circ$ resulting in an additional effective attractive contact interaction with $g^{\rm DDI}_{\rm 1D}=-5.4$. The corresponding values of $\gamma_T$ in the presence (absence) of the intertube DDI are (i) $\gamma_T=2.2$ (2.1), (ii) $4.5$ (4.5), and (iii) $8.6$ (8.5).

\begin{figure}[!t]
    \includegraphics[width=0.99\columnwidth]{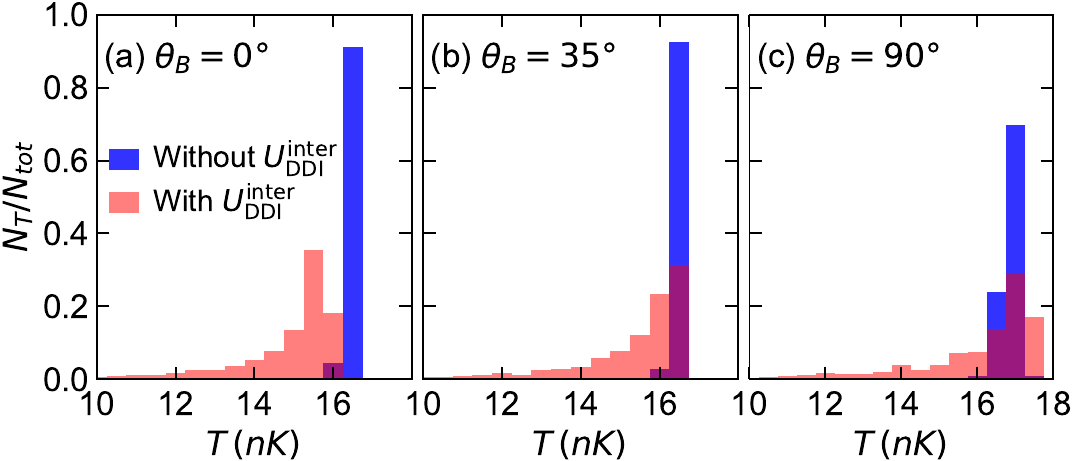}
    \vspace{-0.2cm}
    \caption{Temperature distribution for different values of $\theta_B$. Here $N_T/N_{\rm tot}$ is plotted for initial states with $g^{\rm vdW}_{\rm 1D}=8.5$ $\frac{\hbar^2}{m}\,\mu \text{m}^{-1}$ and (a) $\theta_B=0^{\circ}$, (b) $\theta_B=35^{\circ}$, and (c) $\theta_B=90^{\circ}$. Experimentally, $\theta_B$ is changed adiabatically (keeping the magnitude of magnetic field $B$ constant) after loading into the deep 2D lattice. In our modeling, we modify $T$ in intervals of $0.5$~nK to find the closest entropy in each tube to that at decoupling.}
    \label{fig:temperaturetheta}
\end{figure}

In Fig.~\ref{fig:temperaturetheta} we show our results for the distribution of temperatures after the rotation of the magnetic field. They are qualitatively similar to those reported for $\theta_B=55^{\circ}$. In the presence of the intertube DDI correction, the temperatures are slightly lower and more widely distributed than in its absence. Changing $\theta_B$ slightly shifts those distributions resulting in an average temperature that increases slightly with increasing $\theta_B$.

\begin{figure}[!b]
    \includegraphics[width=0.99\columnwidth]{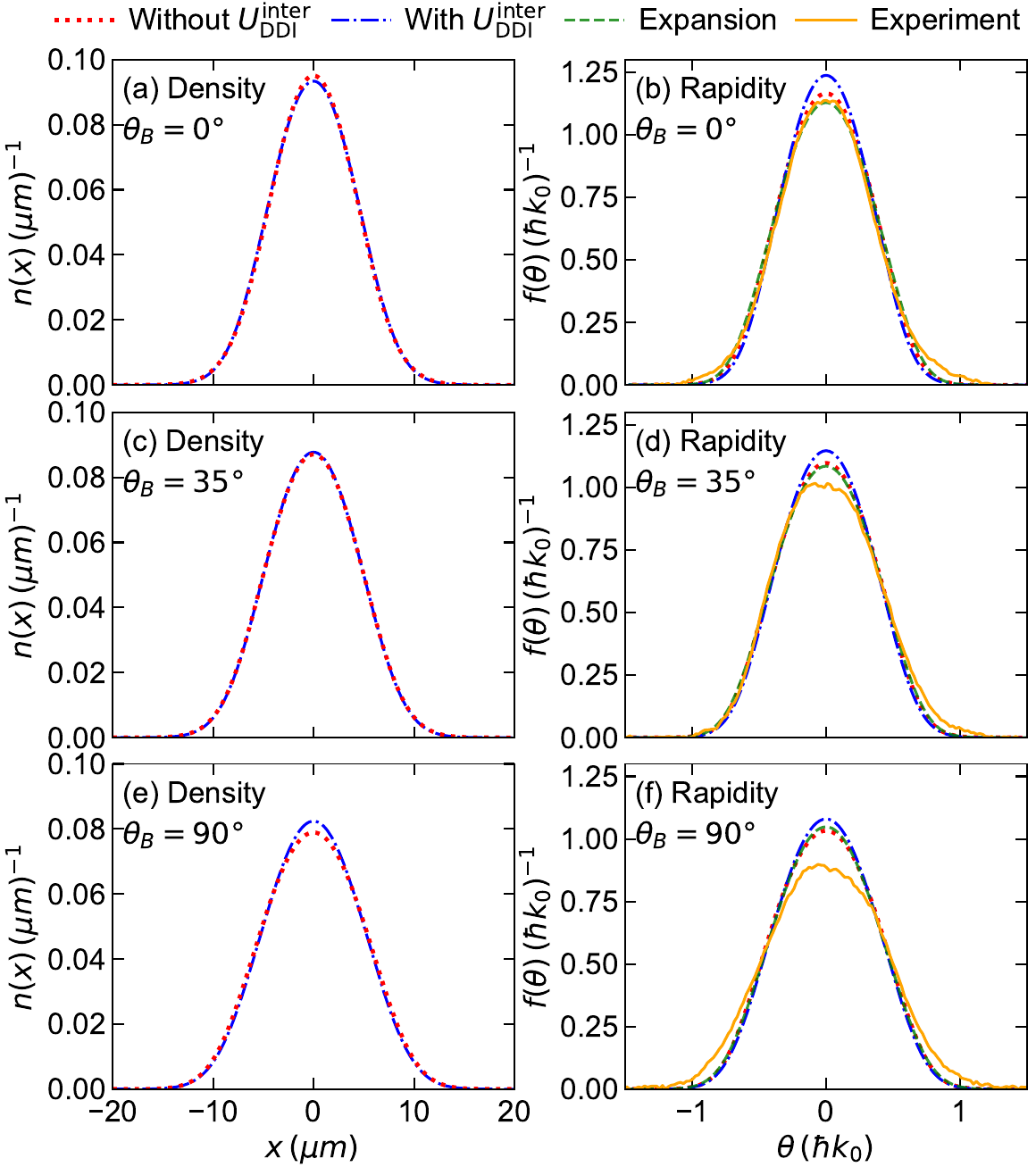}
    \vspace{-0.2cm}
    \caption{Density and rapidity distributions for different values of $\theta_B$. (a), (c), and (e) Density and (b), (d), and (f) rapidity distributions for (a) and (b) $\theta_B=0^{\circ}$, (c) and (d) $35^{\circ}$, and (e) and (f) $90^{\circ}$.}
    \label{fig:rapiditytheta}
\end{figure}

In Figs.~\ref{fig:rapiditytheta}(a),~\ref{fig:rapiditytheta}(c), and~\ref{fig:rapiditytheta}(e) we plot the density distribution without and with the intertube DDI correction for $\theta_B=0^{\circ}$, $35^{\circ}$, and $90^{\circ}$, respectively. The effect of the intertube DDI is small in all cases. As we mentioned earlier, the intertube DDI produces two competing effects on the density distribution: It tends to broaden the density distribution due to antitrapping, while it tends to narrow the density distribution by lowering the temperature of the tubes. The competition between those two effects results in a density distribution in the presence of the intertube DDI that is slightly broader for $\theta_B=0^{\circ}$ [Fig.~\ref{fig:rapiditytheta}(a)], nearly equal for $\theta_B=35^{\circ}$ [Fig.~\ref{fig:rapiditytheta}(c)], and slightly narrower for $\theta_B=90^{\circ}$ [Fig.~\ref{fig:rapiditytheta}(e)] than those in the absence of the intertube DDI.

In Figs.~\ref{fig:rapiditytheta}(b),~\ref{fig:rapiditytheta}(d), and~\ref{fig:rapiditytheta}(f) we plot the rapidity distributions with and without the intertube DDI correction for $\theta_B=0^{\circ}$, $35^{\circ}$, and $90^{\circ}$, respectively. The intertube DDI can be seen to always result in a narrowing of the rapidity distribution with respect to that in the absence of the DDI, and this narrowing effect becomes more prominent as $\theta_B$ decreases. During the expansion, all the rapidity distributions broaden and again they become very similar to those in the absence of the intertube DDI. Overall, decreasing $\theta_B$ results in narrower rapidity distributions both in the modeling and in the experimentally measured rapidity distribution. This feature is properly captured by our modeling, though the peak heights continue to not be accurately simulated. In addition, we note that the agreement between the results of our modeling and the experimental results improve as $\theta_B$ decreases. There is no {\it a priori reason} for this to be the case, so we expect the observed improvement in the agreement to be the result of a cancellation of effects that were not accounted for in our modeling.

\section{Summary and outlook}\label{sec:summary}

We systematically studied the effects of the intertube DDI in recent experiments with arrays of nearly integrable 1D dipolar Bose gases of $^{162}$Dy atoms. The leading-order effect of the intertube DDI is accounted for by treating it as a self-consistent modification to the 1D confining potentials, both during the state preparation and during the rapidity measurements. Our results suggest that this correction changes only marginally the prediction for the measured rapidity distribution. The rapidity distributions obtained by adding the intertube DDI correction are very close to those obtained in its absence because of the nearly perfect cancellation of the effect of the intertube DDI in the initial state with that during the expansion. In the former it leads to a narrowing of the distributions, while in the latter it leads to a broadening, and those two effects are coincidentally nearly identical in our setup. We also found that the intertube DDI slightly cools the 1D gases by generating an effective antitrap. The cooling is more pronounced in the tubes with lower atomic filling lying away from the center of the 2D array of 1D gases.

Having found that the disagreements between the simulations and the experimental results in Ref.~\cite{li_zhang_23} are unlikely to be related to the intertube DDI, we strongly suspect that they are the result of nonthermal effects (due to the near-integrability of the 1D gases) after the tubes decouple. In future work, we plan to explore how to better model the decoupling of the tubes into independent 1D gases. There is no reason for all the tubes to decouple at the same lattice depth $U^*_{\rm 2D}$, and different decoupling depths depending on the filling and location of the tubes may result in a distribution of temperatures at decoupling. Since we found our modeling to be sensitive to the temperature $T^*_{\rm 2D}$ at decoupling, accounting for a distribution of temperatures $T^*_{\rm 2D}$ should provide a better description of the experiments. Also, after decoupling, the 1D gases are nearly integrable, so we will explore descriptions of the ensuing adiabatic processes that take place in the experiments without assuming that each 1D gas is in thermal equilibrium throughout those processes.

\begin{acknowledgments}
Y.Z.~acknowledges support from the Dodge Family Postdoc Fellowship at the University of Oklahoma, M.R.~acknowledges support from the NSF (Grant No.~PHY2309146) and B.L.L.~acknowledges support from the NSF (Grant No.~PHY2308540) and AFOSR (Grant No.~FA9550-22-1-0366).
\end{acknowledgments}

\appendix

\section*{Appendix: Rounding the number of particles and the effect of tubes with very few particles}

\begin{figure}[!b]
    \includegraphics[width=0.98\columnwidth]{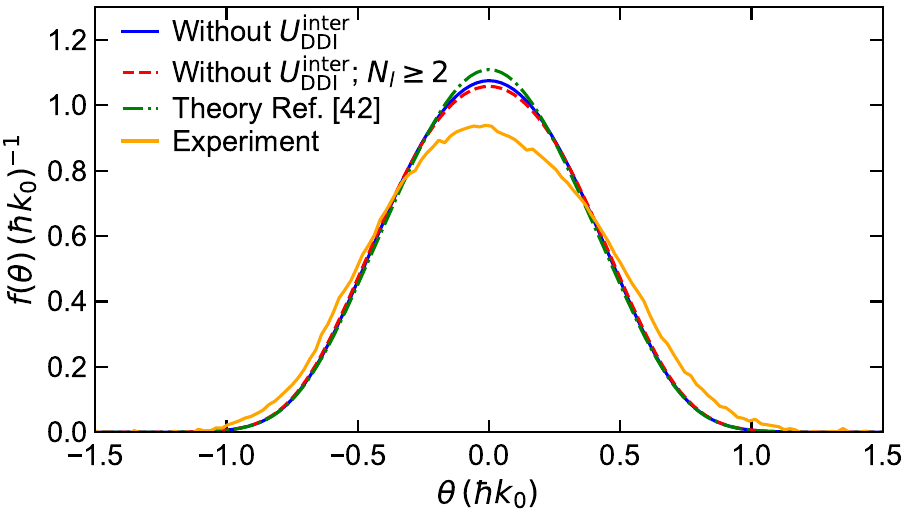}
    \vspace{-0.1cm}
    \caption{Rapidity distributions from this work and from Ref.~\cite{li_zhang_23}. Rapidity distributions for $\theta_B=55^\circ$ and $g^{\rm vdW}_{\rm 1D}=8.5$ $\frac{\hbar^2}{m}\,\mu \text{m}^{-1}$ at the end of state preparation in the absence of the intertube DDI. We show results for the average over all tubes (blue solid line) and over tubes with $N_\ell\geq2$ (red dashed line) obtained in this work, and from the modeling used in Ref.~\cite{li_zhang_23} in which the number of atoms in each tube was rounded to the closest integer (green dash-dotted line). The experimental rapidity distribution from Ref.~\cite{li_zhang_23} is shown as an orange solid line.}
    \label{figapp:rapiditycompare}
\end{figure}

In the modeling in Ref.~\cite{li_zhang_23}, we rounded the number of particles in each tube to the closest integer. This simplification was necessary because the quantum Monte Carlo simulations of the momentum distribution are computationally demanding and therefore it was not possible to simulate thousands of tubes with slightly different fillings. Because in this work we focus on the density and rapidity distributions, which are obtained using the TBA and the LDA, we can carry out the calculations without the rounding. In Fig.~\ref{figapp:rapiditycompare} we show the rapidity distributions obtained in Ref.~\cite{li_zhang_23} and now without rounding (both ignoring the intertube DDI) for the state after fully loading the atoms into the 2D lattice, as described in Sec.~\ref{sec:initialstate}. One can see that, because of the rounding, the rapidity distribution from Ref.~\cite{li_zhang_23} is slightly narrower and further away from the experimentally measured distribution. In consequence, in this work we carry out all our calculations without rounding the number of atoms in the tubes.

On the other hand, because of the self-consistent nature of the calculations carried out in this work to account for the intertube DDI, it is very challenging to include the thousands of tubes that have less than two atoms. As mentioned in the main text, those account for less than 5\% of the atoms. In Fig.~\ref{figapp:rapiditycompare} we show the normalized rapidity distribution obtained after ignoring those tubes. It is very close to that in which those tubes are included (rounding to the closest integer has a stronger effect). Since our main goal is to contrast the results obtained without and with the intertube DDI, we remove those tubes from the calculations (without and with intertube DDI) used to obtain the results reported in Sec.~\ref{sec:finalstate} and in following sections.

\newpage
 
\bibliography{reference}
\end{document}